# Approaches for Modeling of the Complex Heat and Fluid Flows and Another Physical Processes of Fractal Nature

**Ivan V. Kazachkov** [1,2]

[1] *Dept of Information technology and data analysis,
Nizhyn Gogol state university, Grafs'ka, 2, Ukraine, 16600;*
[2] *Dept of Energy Technology, Royal Institute of Technology, Sweden,
Email: ivan.kazachkov@energy.kth.se*

***Abstract.*** *The paper deals with the fundamental problem of a modeling of the physical, in particular, thermal hydraulic processes, in various media of fractal structure of the natural, technological and technical systems and devices. The examples of a few complex systems (mainly, the physical processes) are shown as the different fractal structures, e.g. the ones obtained by cooling and solidification of the multiphase mixtures. Some materials reveal the structural features having a significant influence on the behaviors of these materials. The thermal hydraulic processes which are going during a melts' cooling, with the further water and steam flow through a porous or a channeled media, are presenting the characteristic examples. The best models of the thermal hydraulic and other physical processes in such fractal media should be the ones based on the fractional differential equations. An order of the fractional derivatives in time and space may change in the process. For example, by a cooling of a melt, with a change of the fractal properties of the system, it is going as follows: first, the vapor-water-melt mixture is interacting, then a formation of the solid structure is performing due to solidification of a melt, and a vapor flows through the permeable structure developed in a process. The dynamically changing fractal systems (dynamic fractals) must be modeled by the corresponding equations changing according to the evolving process. Therefore, it is not surprising that so far some of such tasks have not been properly resolved even in a simplified formulation. There are many similar problems in the modern physics, technology, etc., which are discussed in the paper.*

## 1. The vitality of the problem

The model of a continuous medium (continua), having no correspondence in the real nature, works well enough in the tasks at the macro level, since it provides a continuous space, which employs the powerful theories of the functions and the integral-differential calculus. We do not even have instruments for recognizing the immaturity of the continua hypothesis (violation of the hypothesis of continua) at the macro level: any gas, liquid, or solid in all practice is a continuum indeed. We do not record any discontinuity in these media. But in fact, our world is not continuous starting with the atom of any substance that contains only a very small part of the volume occupied by the nucleus of protons and neutrons, while the main part is a cloud of moving electrons by their orbits. Possibly the nucleus of an atom, in turn, is also empty to a large extent. And only thanks to the electromagnetic and nuclear forces, this entire void appears in the macro world as a solid, liquid or gaseous matter. Thus, the material world is, in fact, rather a macromvage of the energy processes, a manifestation of the energy. At the macro level, we have to distinguish these concepts considerably and even to consider the matter as the primary, not knowing what it is, and the energy - a derivative of matter that produces it under certain conditions. But our fundamental problem, which is posed for the subject of this work, is much narrower, and it consists in the fact that such a pleasant and a simple model of a continuum (continuous space) does not work in many macro-level tasks.

Examples of the tasks that are in a complete discrepancy with the continuum model at the macro level seem to be all the tasks of the multiphase and multi-component systems, where the different phases and components distributed in a space have their own boundaries or are mixed in a one volume if they are the interpenetrating and the inter-soluble liquids or gases (alcohol-water, water-air, for example). The insoluble immiscible liquids behave as the multiphase media (water-oil with the dynamic moving surfaces as an example of a liquid-dispersed liquid, or liquid with solid particles - as an example of liquid-dispersed solid particles with non-deformed boundaries of the phase transition). In general, the multi-phase multicomponent system can have many fragments of the different sizes and shapes for the given phases and components, with the deformed and undeformed surfaces of a partition of those media in a space. In the volume of each of them, the continuum hypothesis operates, and the conditions for the interaction of the contacting media on the boundary surfaces for each of them are there. Since they are, as a rule, very large, then putting the same number of the conjugate boundary problems for each of them is a very difficult problem that can not be solved. This is especially complex due to the fact that the boundary conditions must be stated at the moving boundaries that are constantly deformed in accordance with the processes occurring in such contacting continua. These processes, in turn, have an influence of the boundary conditions on the surfaces of contacting continua. The boundary-value problems for the regions with the moving boundaries are not yet resolved theoretically even for two simple contacting regions, but here can be many of them.

For example, in the problem of cooling the molten corium during severe accidents at nuclear power plants in the system of passive protection from accidents, the melt should be kept in a controlled thermal state [1-6]. One of such systems of the European EPR reactor includes a sub-reactor pool in which water in the event of an accident penetrates to the melt layer of the corium from below through the nozzles located below the plane of the bottom (after removing the thin layer of sacrificial concrete), intensively evaporates at the inlet to the high-temperature melt, mixes with the melt throughout the volume, taking the warmth away from it and removing it outside the container with a vapor [6]. The melt is gradually cooled. Local point cooling of the melt to the point of freezing appears in different places. And solidification of the melt begins in these individual locations. There is a melt, the solid corium's particles of varying sizes distributed by a volume. A vapor is stirred, in some places, depending on the local thermohydraulic parameters. The particles can melt, while in other locations the new particles can be formed where the melt temperature is dropped below the temperature of the corium's melting. Such process continues until the solid particles be formed so much that they begin to form agglomerates, which are dynamically formed and destroyed in some places due to an arrival of other portions of the hotter liquid or due to the mechanical action of the intense jets of the formed vapor; and formed again due to the arrival of other portions of the volume of colder coolant, and so on. Finally, a porous system with a different structure, different permeability is formed, which later on needs to be kept in a controlled state; otherwise there may be created some areas of inadequate cooling, where a corium melt will be formed again that will further warm up as a result of radioactive decay with the internal heat generation.

Such tasks at the macro level have a pronounced fractal nature, and therefore their description within the framework of a continua makes no sense. Despite the great efforts in the creation of the mechanics of multiphase multicomponent systems based on various modified models of continuum media, since the 1950's, and significant advances in a solution of many important problems of this direction, relatively simple from the above point of view [7, 8], all these theories are deadlocked. Obviously, it is necessary to change the initial principles, the basic concepts. The so-called multi-speed and interpenetrating continuities do not reflect the completely fractal nature of the systems. In this case, the geometry, as well as the processes themselves should be considered as fractal by nature. Moreover, all the prerequisites were created for this: the basis of fractal geometry [9] and the integrodifferential analysis of the fractional order for the description of processes in fractal structures [10-32]. Taking into account the above, this article is devoted to consideration of the problem of creating a methodology for the modeling of thermal hydraulic and other physical processes in the systems of fractal nature and derivation of the generalized fractional order equations for such processes occurring in the fractal spaces and having fractal nature by space and/or time.

## 2. Statement of the problem by modeling of complex systems of fractal nature

Let's give one example of objects of fractal nature important for study due to the needs of practice, for which the classical methods of continuum mechanics are completely unsuitable, except for the integral simple estimates. Thus, one of the passive systems for protection against severe accidents at the NPP is the system of rapid cooling of the corium proposed in the European reactor EPR [33-37], presented in a simplified form in Fig. 1 [38]. The EPR with the electric power of 1600 MW is developed by the French firm Framatome and the German Siemens-KWU (now Framatome ANP) based on the French N4 and the German PWR "Konvoi". The concept of EPR safety reflects the general trend of the philosophy of serious accidents. During a long removal of the excess heat from the container, the EPR uses a special cooling system for the container. As shown in Fig. 1, the water enters the system from the internal refillable water tank, serves the cooling system of the container through an external heat exchanger and returns water to the container [38]. This excludes an additional reduction in the reliability of the system due to an exclusion of its external devices. The system for removing a heat from the container has 2 schemes and 2 principles of operation.

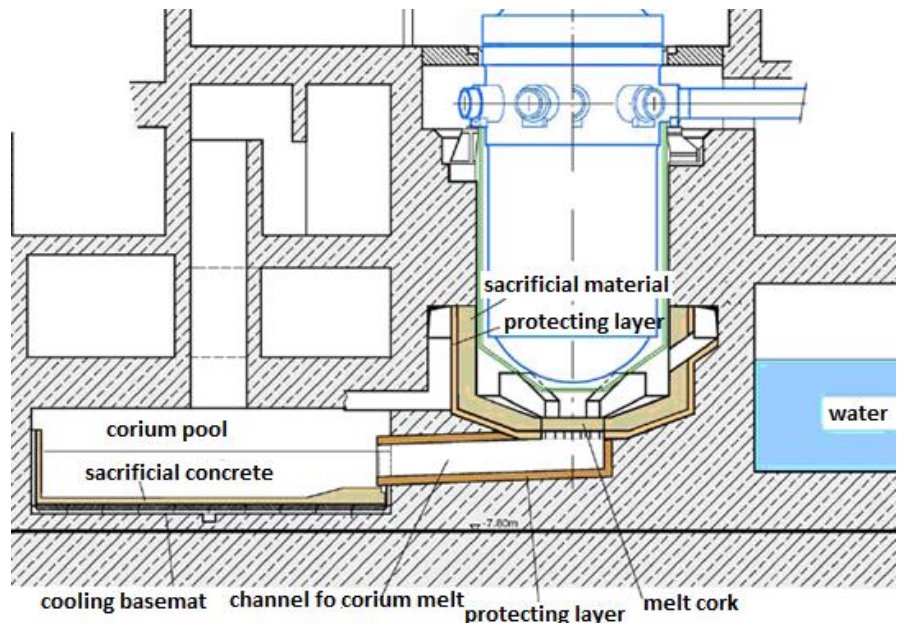

Fig. 1 The main components of the stabilization system for corium melt in EPR [38]

Many works were devoted to a study of the complex problems concerning the corium melt's interaction with a coolant in a number of the conditions and scenarios of the hypothetic severe NPP accidents: at the Karlsruhe Research Center, the Royal Institute of Technology (KTH), and the other research centers and universities [39-45]. Examples of the results obtained at KTH are shown in Figs 2-5 for a modeling of water cooling the high-temperature melts of the calcium and vanadium oxides (CaO + $WO_3$) [42, 45]. The initial temperature of the melt was 1250 °C. The coolant (water) was fed through 5 nozzles at the bottom of the cooling pool. A study of the resulting porous material showed its high porosity (approximately 38%) evenly distributed in a volume (Fig. 2). In the other experiment (MnO + $TiO_2$) at the temperature 1350 °C (Fig. 3), the porosity of a solidified material was about 47% evenly distributed in a volume.

The porous structures obtained by Domenico Paladini in his work on the licentiate's dissertation, with the other materials, are presented in Figs 4, 5. They show an influence of another parameters and conditions, where the more viscous modeling melts do not mix as intensively as in the cases in Figs 2, 3 [42, 43, 45]. After the melt was frozen, a permeable material of a lower penetration and a completely different structure was obtained. In Fig. 4 [41, 44, 45] there are observed some separate channels, where the coolant accidentally choked the way through a fairly viscous melt, whereas in Fig. 5 [41, 44, 45] the structure of the resulting material is seen as porous but not very well permeable and uniform in a volume. There are also the pores of different sizes and the separate channels. Both cases are worse than the ones shown in Figs 2, 3; they correspond to a poor cooling of the melt due to a low permeability of the material.

The following cases of the different fractal structures of the cooled materials have shown that the structural features have a significant influence on the behaviors of the thermal hydraulic processes during cooling of the materials with water and steam flowing through the porous, channeled, or combined media. The fractal structures of the permeable media are the best models to their nature, both in the frozen state after cooling, and in the cooling stage from the beginning of the coolant penetration through the melt layer to a cooling of the permeable solid corium or the other material obtained after solidification of a melt. Therefore, the best model of thermal hydraulic and other physical processes in such a fractal media should be based on the differential equations in fractional derivatives, in which the order of fractional derivatives in time and space changes in the process of cooling the melt, with the change of fractal properties of the system. So it is obvious that we have cases of dynamically changing fractal systems, that is, dynamic fractals. This means that the equations describing such sys-tems are constantly changing. Initially, there is an intense mixing of the melt, water, and steam, and the phases and components of the liquids and gases (vapor) in a volume are fractionally distributed.

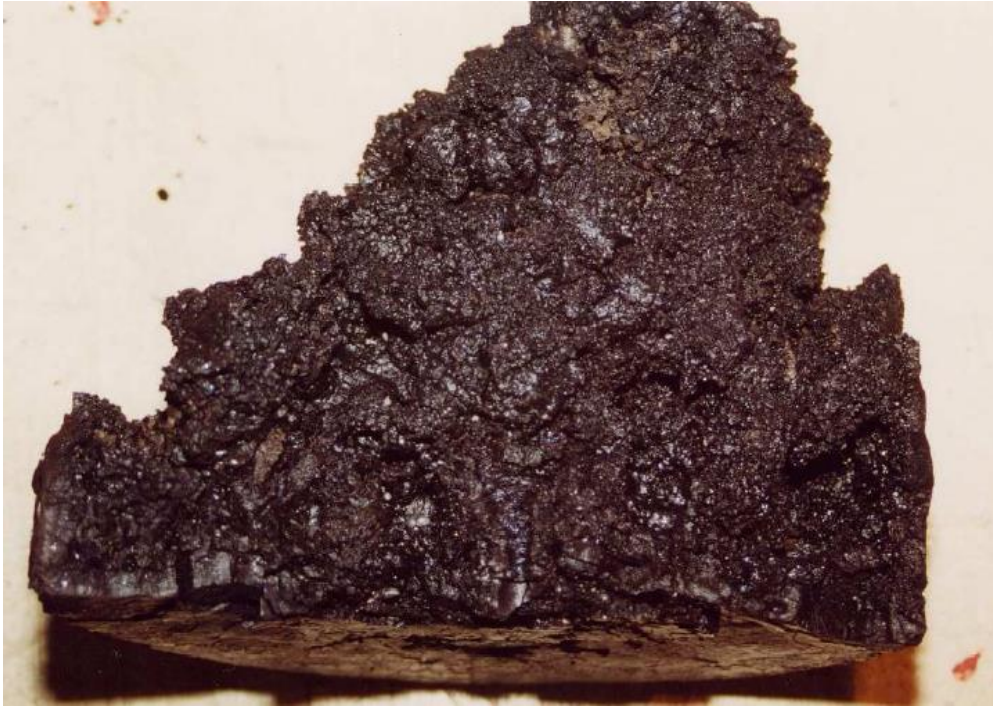

Fig. 2 Porous material $CaO+WO_3$ obtained at KTH by water cooling of the melt

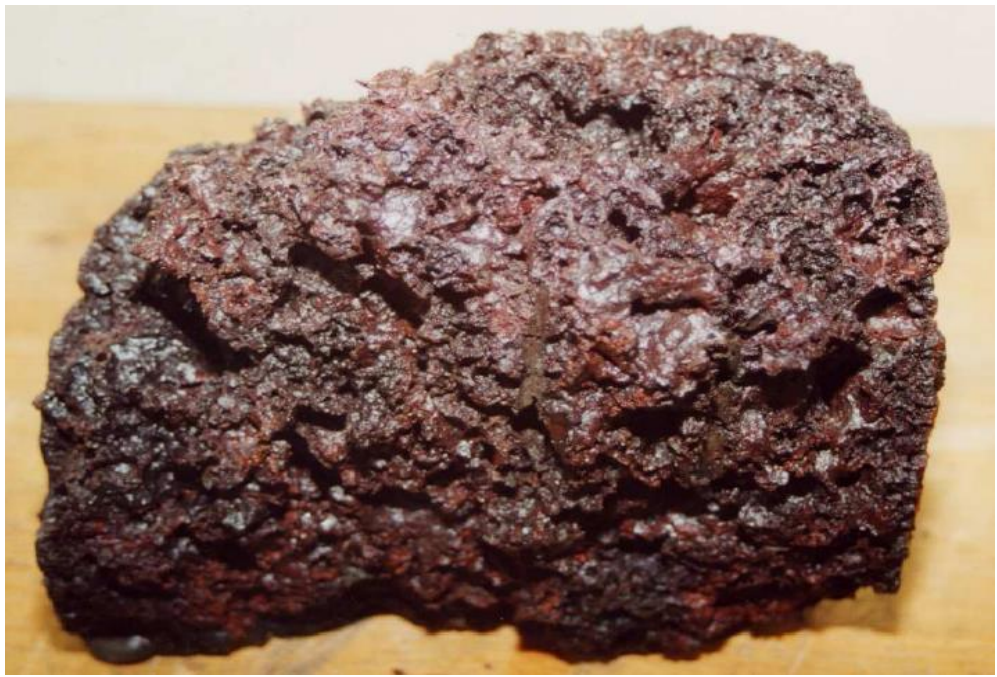

Fig. 3 Porous material $MnO+TiO_2$ obtained at KTH by water cooling of the melt

When the solid particles have been formed, then the other structure was created. And the beginning of a formation of the skeleton of the porous material presented a quite different new structure. And so on. This is only a fragmentary allocation of the basic fractal structures. In fact, the process is dynamic and much more complex. In addition, the mathematical models of physical processes in such systems must also be constantly changing. Therefore, it is not surprising that this task has not been resolved so far even in a simplified formulation. There are many similar problems in the modern physics, technique and technology. More details are discussed below.

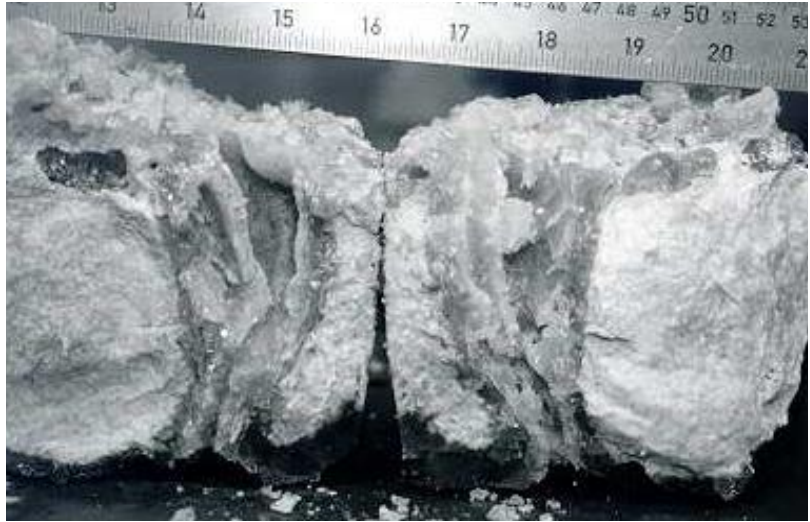

Fig. 4 Porous material obtained by Dr. D. Paladino in cooling of the high viscous salt by water

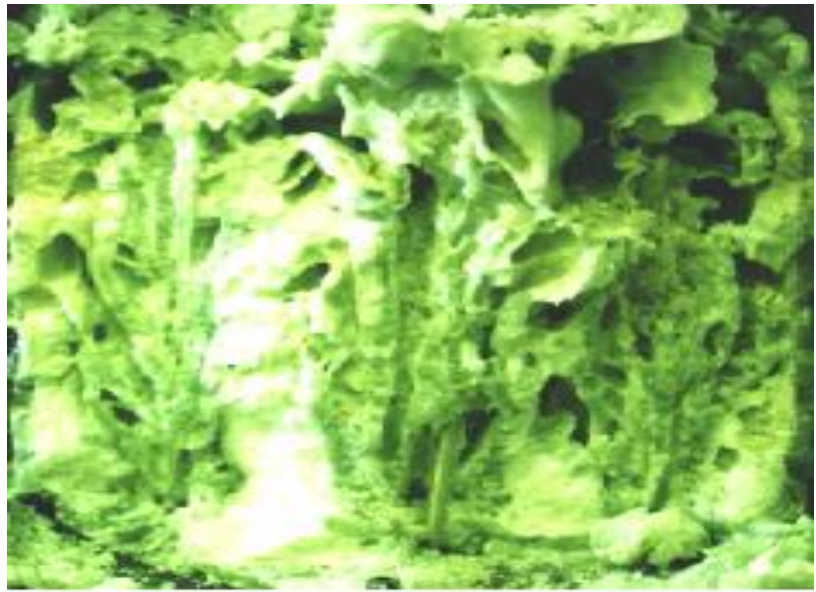

Fig. 5 Porous materials obtained by Dr. D. Paladino in cooling of the middle-viscous salt by water

## 3. Development of the physical-mathematical models for thermal-hydraulic processes in fractal systems

The equations with derivatives of the fractional order describe the different models of abnormal diffusion, which arise in the amorphous, colloidal and porous materials, in the biological systems, polymers, random and sparse media [10-14]. For example, the authors [10, 11] have obtained a solution of the one-dimensional heat conductivity equation determined on a segment [0,1] containing the derivatives of a fractional order in a time and in a spatial variable, which were considered by the definition of Caputo. The influence of the thermal pulse, which was given on the boundary, on the temperature field was investigated. The results of calculations showed the transition from the pure diffusion heat conduction to the wave heat conduction process. Thus, the fractional differential equation of a heat conductivity was considered in [10] in the simplest canonical form [22, 31]:

$$\frac{\partial^{\alpha} T}{\partial t^{\alpha}} = \frac{\partial^{\beta} T}{\partial x^{\beta}}, \qquad (1)$$

where $0 < \alpha \leq 2$, $1 < \beta \leq 2$, $0 \leq x \leq 1$, $t > 0$, $T(x,t)$ is the temperature, $\frac{\partial^{\alpha}}{\partial t^{\alpha}}$, $\frac{\partial^{\beta}}{\partial x^{\beta}}$ are the operators of the fractional differentiation by Caputo [27, 28]. The following uniform initial and boundary conditions have been considered:

$$t = 0, \quad T = 0, \quad 0 < \alpha \leq 1; \qquad t = 0, \quad T = \frac{\partial T}{\partial x} = 0, \quad 1 < \alpha \leq 2; \qquad (2)$$

$$x = 0, \quad T = H(t); \qquad x = 1, \quad \frac{\partial T}{\partial x} = 0. \qquad (3)$$

Here $H(t)$ is the Heaviside function.

If the Laplace transformation for the fractional derivative by Caputo [22] is applied in the equation (1) with the boundary conditions (3), accounting the initial conditions (2), then the boundary problem for the equation in the transforms is got. After obtaining a solution of the equation in transforms, the inverse Laplace transformation [46] allows determining the temperature *T(x,t)* [10]. An example of a solution is given in Fig. 6 for x = 0,5 (middle of the considered interval). The result shows that an influence of the parameter α (it reflects a fractal nature of the investigated processes dynamics in time) is not just quantitative but the qualitative one, which is the most surprising. The type of solution is changing with this parameter from a monotonous one to the wavy process. An influence of the parameter β is mainly quantitative (it reflects a fractal nature of the investigated process dynamics by a special variable).

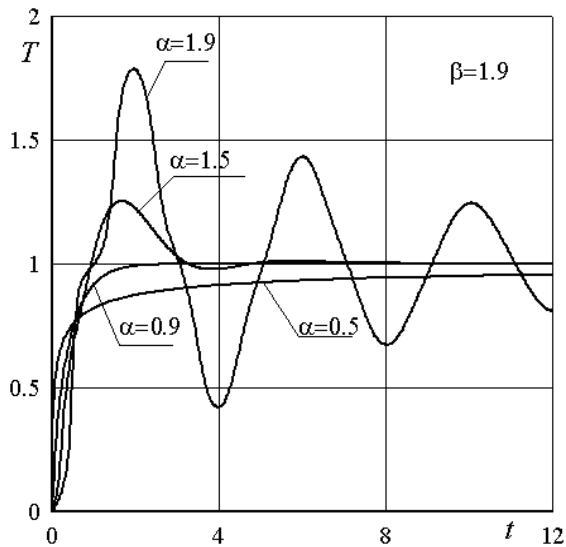

Fig. 6 The temperature depending on time at the point *x*=0,5 m by different values α

In the paper [10], on the basis of the obtained solution, the process of fractal thermal conductivity, which is carried out due to the sharp temperature heating of the boundary of the region, was investigated. Only the authors made a small significant physical error in one place, arguing that at x = 0 m the boundary condition is fulfilled: $T = 1^0 K$ (!). At temperatures close to absolute zero there is no conventional heat conductivity, the nature of which is a chaotic motion of molecules, which is nearly absent by absolute zero temperature. And the equations of heat conductivity in this area of temperature are unfair. Of course, in order to eliminate this misunderstanding, it would suffice to indicate that the value of *T* in equation (1) is the temperature difference in the region relative to some characteristic temperature of the system. Then there would not be such an error and there would be no talk of the absolute zero temperature. There is also a misunderstanding with the fractional dimensions of time and linear dimensions, so it is better to consider the equation in a dimensionless form.

The solution continuously depends on the parameters α, β and changes from the solution of the pure diffusion equation (α = 1, β = 2) to the solution of the wave equation (α = 2, β = 2). Thus, at α = 0.5, β = 1.9 there is a slow process of a heat conduction: the temperature reaches the equilibrium value in about 12 seconds, while according to the classical heat conduction equation (α = 1) $t \approx 2$ s. At α =

1.9, the process of thermal conductivity has a wavy character; at the time $t = 1.5$ s a reflection from the boundary occurs. The fractional parameter β takes into account the spatial fractal features of the system, so the temperature varies in different ways at the different values of β, but the thermodynamic equilibrium of the system occurs at approximately the same moment of time $t = 2$ s.

The fractional derivative is a nonlocal characteristic of a function, as far as it depends on a behavior of a function not only in the vicinity of the observation point $x$ but also on the values it has acquired over the entire interval $(a, x)$ or $(x, b)$. That is why it can be applied to the description of the fractal processes. The requirement of continuity of the space and of the processes is removed.

The hereditary probabilistic processes whose rate of change in density depends on the density values at the previous moments of time [14] are conveniently described by the equations containing a fractional derivative in time, the order of which is determined by the value of α, which characterizes the topology of the given set [15]. Two types of an abnormal transfer are possible: subdiffusion and superdiffusion. The slow diffusion (subdiffusion) is characterized by the values of the index $0 < α < 1$, the fast diffusion (super diffusion) is characterized by the values of the index $1 < α < 2$, and at the boundary of these two regions, with $α = 1$, there is a normal diffusion process described by the Fick's law [16]. The connection between anomalous heat conductivity and diffusion in the one-dimensional systems is established, for example, in [17]. As well as the authors of the paper [10] pointed out, in the recent years, the researchers have received a considerable interest to the fractional diffusion equation (thermal conductivity), the reasons of which we have revealed somewhat in our studies of the above-mentioned multiphase systems.

The interesting solutions of the diffusion-wave equations were obtained during the recent decades. For example, F. Mainardi obtained the fundamental solution of the one-dimensional diffusion-wave equation [18]. W.R. Schneider and W. Wyss [19] and Y. Fujita [20] solved the Cauchy problem for the fractional diffusion and wave equations using the method of transformation to the integral-differential equation. A. Hanyga [21] used a time-varying diffusion equation to study the propagation of the elastic waves in a viscoelastic medium. And Y.Z. Povstenko gave a solution to the diffusion and thermoelasticity equations with a fractional derivative in time [22-24] and considered [25] a two-dimensional diffusion-wave equation with the fractional derivative of Caputo [26-28]. By the method of the integral transformations, the Dirichlet and Neumann boundary problems were investigated and the calculations ware presented [29, 30]. Sh. Momani [21] applied the Adomian decomposition method for solving the fractional differential equation with a spatial derivative of Caputo. Information on the origin of the fractional derivatives of the Riemann-Liouville and Caputo and their application can be found in the review articles [16, 22], where a general solution of the linear differential equation with a fractional derivative of Caputo was done too [22].

During the last decades, the class of differential equations in fractional derivatives has been successfully applied to the simulation of complex processes and systems of various nature, in particular, physical and physical-chemical. For example, fractional diffusion and wave processes. In addition to the above works, one can indicate different universal electromagnetic, acoustic, mechanical and other processes described by the fractional diffusion-wave equations [47-51], which correspond more accurately to the fractal nature of processes than the classical equations constructed on the hypothesis of continuous space and continuous processes.

## 4. The parameters of fractional differential and fractional integral equations

The determination of the parameters, which state an order of the fractional derivatives, is one of the most complex problems for practical application of the integral-differential equations of fractional order for mathematical modeling of the diverse complex process At the above-considered examples there were the values α, β, which show the nonlocal properties in a time and space, respectively. The derivative of the fractional order of functions in time, as a nonlocal characteristic of a function, which depends on the behavior of the function not only at the current time, but also on the values acquired by it over the entire interval of time to the present moment (the previous history of system development, process), as well as future behavior functions (biases, predictions, orientation of the system's development into future indicators) contains, in essence, all information on the peculiarities of development in time. And this degree of non-localization in time is given by the parameter α, the effect of which can be investigated similar to the above, but it is not clear how to set the real value of this parameter for a particular system or process.

Similarly, the fractional derivative of a function of a spatial coordinate in the vicinity of a point of observation x also has a nonlocal character, now in space, and depends on the values it has acquired over the entire interval (a, x) or (x, b). The degree of this nonlocality of a function represented by the parameter β can be considered as a characteristic of the fractal properties of the system under consideration. But how to determine the exact value of the parameter β for a particular system is not clear. For fractal objects in the dimension of dimension, their two fractal dimensions can vary in the range from 1 to 2, so in the first approximation we can identify β with the fractal dimension of the system, which can be calculated for the known object topology [52], but for one-dimensional or The three-dimensional heat equation is unclear how to determine β. In addition, one and the same fractal dimension of objects does not mean their identical structure, which can affect the processes of heat conductivity or other simulated processes more strongly than the fractal dimension. Therefore, the question of determining the parameter β remains unclear. The only positive fact is that, as shown in [10] and in many other papers, the effect of β is mainly quantitative, whereas the effect of α is both quantitative and qualitative.

The nonlocal behavior of the various systems in time, studied on many examples of systems from a miniature to cosmic scale, from living and inanimate nature [53], showed that even solutions of the simplest ordinary differential equations of a development, with the deviated arguments, have the complex features and critical modes. Thus, the study of the phenomena of critical regimes in the developmental equations, with the delayed and advanced arguments in time, showed that the strategy of a stable development of the system, taking into account the delayed and forecasting arguments, should be lump-continuous. And it should be regulated in the interval from the critical curve of the development strategy with the advance arguments to the critical curve of the development strategy with a delay in time. Fig. 7 shows the possible strategies for the stable development of the system [54]:

$$\frac{dx}{dt} = kx(t \mp \tau) , \qquad (4)$$

where $k$ is a coefficient and $\tau$ is a time shift (time delay or time forecast)

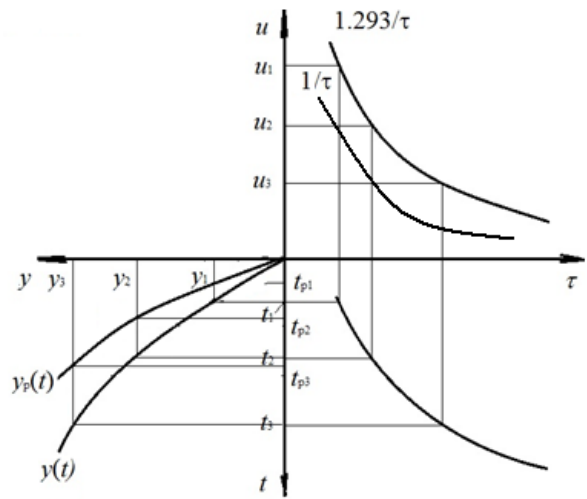

Fig. 7 The schematic prognosis of the stable development of the system with account of the critical levels and strategies with the delay and forecasting in time

As seen from Fig. 7, the possible strategies in development are going below the curve $u=1.293/\tau$ depicted in the first quadrant (the boundary of the area of the stable development) for the development equation (4) with a delay $\tau$ in time [54]. For the development equation with an advance $\tau$ in time, the possible strategies in development must be above the critical curve $u=1/\tau$. Therefore, the optimal strategy for development can be predicted as follows. The growth of a system begins with the desired tempo $u_1$ set at the beginning. Then such tempo can be only kept until the moment of time when the increasing delay of the system at time $t=t_1$ becomes critical (the line $u=u_1$ crosses the critical curve $u=1.293/\tau$) at the point $\tau_1(t_1)$. Afterwards, a further growth of the system with a given increasing rate is impossible under the condition of a further growth of the delay. It is necessary to reduce the growth rate of the system, for example, to some value $u=u_2$, after which there is an additional resource of the system in relation to a growth of the time delay, up to crossing the critical curve at the lower level of the growth. But not below the critical curve $u=1/\tau$, which corresponds to a development of the system with the allowance for the advance in time (the time forecasting terms).

In practice, the system can be more complex and its mathematical model can be, for example, the equation array with a number of different deviated arguments [55-59]. Such systems can have much more peculiarities in their behaviors and have many critical levels in their development. One of them is illustrated in Fig. 8 for the aggregate mathematical model of the potentially hazardous object built by us.

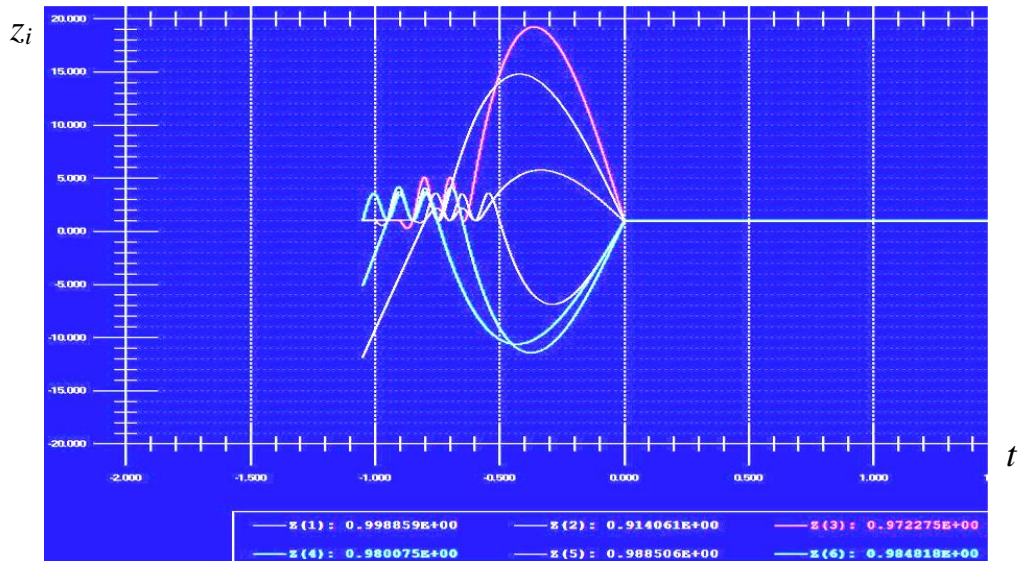

Fig. 8 Functions of a development $z_i$ (calculated, $t>0$) and interpolated «history of development» ($t<0$)

## 5. The Conclusions and further development

We have considered the thermal hydraulic processes and the solidification of the melts in the domains of fractal geometry, which can be met in the problem of the severe accidents' modeling and simulation, as well as in many other tasks too. In such complex case, there are available dynamical fractals of evolutionary geometry and fractal properties in space and time as a concern to the physical processes. The last one was shown on a simple example of the diffusion-wave heat conductivity equation. Thus, in general, in this system, we deal with the multiple dynamical fractals of the changing geometry and of the changing physical properties of a fractal nature.

The other examples from the systems described by the differential equation with deviated arguments have shown that the delays and advance in time cause nonlinear properties and fractal nature even in the comparably simple linear differential equations of a development. Hereditary processes are studied in diverse fields for a long time, and these kind features were revealed for different systems. What is more, in all computer simulations, the approximate difference equations are implemented as the models for the outgoing differential equations, which mean a transforming from the continuity to the discontinuity. This means that all approximate difference equations as the models for differential equations are nonlocal so that any discretization of the outgoing equations cause specific type fractal properties instead of continuity (locality) in the outgoing equations. Thus, the fractal properties are born by all numerical (finite-difference and finite volume) methods.

The specific features underlined here must be accounted. By the modeling of the different systems, we have to elaborate these ideas to reveal some more interesting features connected with the fractal geometries and the fractal nature of the processes. As shown in [60], the increase of fidelity in an approximation of a nonstationary solution by time requires a corresponding increase of the region involved by space. The nonlocal properties by time are directly connected to the nonlocal properties by space so that the structure is of the fractal nature.